\documentclass[%
aps,
pra,
superscriptaddress,
tightenlines,
showpacs,showkeys,
a4paper,
12pt,
longbibliography,
notitlepage
]{revtex4-2}

\usepackage[english]{babel}
\usepackage[utf8]{inputenc}
\usepackage[T2A]{fontenc}

\usepackage{amssymb,amsmath,stmaryrd,array}
\usepackage{graphicx}
\usepackage{multirow,float,subcaption}

\makeatletter
\newcommand{\ah}{\hat{a}}
\newcommand{\Ah}{\hat{a}^\dagger}

\newcommand{\ket}[1]{\ensuremath{|{#1}\rangle}}
\newcommand{\bra}[1]{\ensuremath{\langle{#1}|}}

\newcommand{\avr}[1]{\ensuremath{\langle{#1}\rangle}}

\newcommand{\cnj}[1]{{#1}^{\ast}}
\newcommand{\hcnj}[1]{{#1}^{\dagger}}

\newcommand{\Tr}{\mathop{\rm Tr}\nolimits}

 \newcommand{\ind}[1]{\mathrm{#1}}

\newcommand{\dd}{\mathrm{d}}

 \newcommand{\e}{\mathrm{e}}

\makeatother

\begin{document}
\DeclareGraphicsExtensions{.png,.pdf}
\title{
  Sub-Poissonian light
  in fluctuating thermal-loss bosonic channels
}

\author{Iliya~G.~Stepanov}
\email[Email address: ]{i.stepanov@itmo.ru}
\affiliation{ITMO University, Kronverksky Pr. 49, Saint Petersburg  197101, Russia}

\author{Roman~K.~Goncharov}
\email[Email address: ]{toloroloe@gmail.com}
\affiliation{ITMO University, Kronverksky Pr. 49, Saint Petersburg  197101, Russia}

\author{Alexei D.~Kiselev}
\email[Email address: ]{alexei.d.kiselev@gmail.com}
\affiliation{Laboratory of Quantum Processes and Measurements, ITMO
  University, Kadetskaya Line 3b, Saint Petersburg 199034, Russia} 
\affiliation{Leading Research Center "National Center for Quantum Internet", ITMO University,
  Birzhevaya Line 16, Saint Petersburg 199034, Russia}


\date{\today}

\begin{abstract}
  We study the photon statistics of
  a single-mode sub-Poissonian light
propagating in 
the temperature-loss bosonic channel with fluctuating transmittance  
which can be regarded as a temperature-dependent model of turbulent atmosphere. 
By assuming that the variance of the transmittance
can be expressed in terms of the fluctuation strength parameter
we show that the photon statistics of the light remains sub-Poissonian 
provided the averaged transmittance exceeds its critical value.
The critical transmittance is analytically 
computed as a function of the input states parameters, the temperature
and the fluctuation strength.
The results are applied to
study special cases of the one-mode squeezed states
and the odd optical Shr\"odinger cats.
\end{abstract}
 
 \maketitle

\section{Introduction}
\label{sec:intro}

The nonclassical properties of optical fields
lie at the heart of quantum optics
and
from its very beginning they have been the subject of
numerous intense studies.
There are a number of indicators
introduced to measure quantumness (nonclassicality) of light
such as
negativity of the Wigner function~\cite{Kenfack:jopt:2004},
squeezing~\cite{Teich:qopt:1989,Lvovsky:inbk:2015}
and sub-Poissonian statistics~\cite{Davidovich:rmp:1996}
(see also a recent review on
quantumness 
quantifiers based on Husimi quasiprobability~\cite{Goldberg:avs:2020}).

Note that,
for the squeezing and the sub-Poissonian statistics,
the indicators are formulated in terms of
second-order moments of fluctuations of
the experimentally measured quantities.
For nonclassical fields,
these moments violate certain inequalities. 
For example,
the sub-Poissonian light is indicated when
the Fano factor defined as
the ratio of the photon number variance and the mean photon number
is less than unity.

Aside from its fundamental importance,
light nonclassicality plays a vital role
in quantum metrology~\cite{Tan:avs:2019}.
For sub-Poissonian fields that will be our primary concern,
there are various experimental techniques
used to generate sub-Poissonian
light~\cite{Li:pra:1995,Perina:pra:2021,Iskhakov:ol:2016,Lal:revinst:2019,Samedov:physatom:2021}
and their applications in quantum imaging
are reviewed in~\cite{Berchera:metrol:2019}
(higher-order sub-Poissonian statistics is discussed in~\cite{Erenso:josab:2002,Perina:pra:2017}).
Security analysis of BB84 protocol with sub-Poissonian light sources was performed in~\cite{Waks:pra:2002}.
Inﬂuence of temporal ﬁltering of sub-Poissonian single-photon pulses on the expected secret key
fraction, the quantum bit error ratio, and the tolerable channel losses
is analyzed in~\cite{Kupko:npj:2020}.

It is well known that
continuous variable quantum states of
non-classical light used in
quantum metrology and
quantum communication protocols~\cite{Pirandola:advopt:2020,Kiselev:nano:2022} 
are subject to loss and added noise leading to degradation of non-classicality and quantum correlations.

For free-space communication links~\cite{Peuntinger:prl:2014,Heim:njp:2014,Derkach:njp:2020,Pirandola:prres:2021},
a widely used general theoretical approach to
modeling environment-induced decoherence effects
is based on Gaussian quantum channels with fluctuating parameters.
Specifically,
a pure-loss channel with fluctuating transmittance exemplifies
a popular model that describes the propagation of quantum light
in a turbulent atmosphere
(see a review on propagation of classical eletromagnetic waves through a turbulent atmosphere~\cite{Kravtsov:rpp:1992}).

This model has been extensively used to study
nonclassical properties and quantum correlations
of light propagating in turbulent
atmospheres~\cite{Milonni:joptb:2004,Kiesel:pra:2008,Semenov:pra:2009,Vasylyev:prl:2012}.
In Ref.~\cite{Gumberidze:pra:2016},
Bell inequalities in turbulent atmospheric channels
are explored using
the probability distribution of transmittance (PDT) in the elliptic-beam approximation
with parameter suitable for
the weak to moderate-turbulence channels~\cite{Vasylyev:prl:2016}.
Gaussian entanglement in turbulent atmosphere
and protocol that enables entanglement transfer over arbitrary distances~\cite{Bohmann:pra:2016,Avetisyan:optexp:2016}.
The evolution of higher-order non-classicality and entanglement criteria in
atmospheric fluctuating-loss channels are investigated in~\cite{Bohmann:pra:2017}.
Theory of the classical 
effects associated with geometrical features of light propagating, such as beam wandering, widening
and deflection is developed in~\cite{Vasylyev:pra:2018,Vasylyev:pra:2019}.
In Ref.~\cite{Klen:pra:2023},
the PDT derived by numerical simulations is compared with the analytical results.

In this paper, we
adapt a generalized model of the channel
and use the thermal-loss channel with fluctuating transmittance
to examine how
the temperature effects combined with
the fluctuating losses influence
the sub-Poissonian light.

The paper is structured as follows.
In Sec.~\ref{sec:model} we describe a temperature-loss channel
and the parameters
expressed in terms of the first-order and second-order moments
of the photon number
used to identify sub-Poissonian light fields.
In particular,
we deduce the input-output relation
for the $q$-parameter
introduced as a un-normalized version of the Mandel $Q$-parameter.
In Sec.~\ref{sec:fluct}
this relation is generalized to the case of the temperature-loss channel with
fluctuating transmittance.
After parameterization of
the transmittance variance, it is shown that
the output light will be sub-Poissonian
only if the average transmittance exceeds
its critical value.
In Sec.~\ref{sec:temp}
we
apply the theoretical results to the specials cases of
squeezed states, odd optical cats and Fock states
and study how the critical transmittance
depends on the temperature and the strength of transmittance fluctuations.
Finally, concluding remarks are given in Sec.~\ref{sec:discussion}.

\section{Channel and moments}
\label{sec:model}

We consider a single-mode quantized light
with the annihilation and creation operators,
$\hat{a}$ and $\hcnj{\hat{a}}$,
propagating through a quantum Gaussian channel.
The simplest and widely used
method to describe the channel
is to
introduce additional bosonic mode representing
the degree of freedom of environment
and assume that the interaction between
the light and noise modes
is determined by the channel unitary
$\hat{U}_{\tau}$ giving
the beam splitter transformation
of the form:
\begin{align}
\label{eq:a_evolution}
 \hat{U}^{\dagger}_\tau \hat{a} \hat{U}_\tau\equiv
\hat{a}_\tau= \sqrt{\tau} \hat{a} + \sqrt{1-\tau}\hat{b},
\end{align}
where  $\tau$ is the channel transmittance 
and $\hat{b}$ is the aninihilation operator
of the noise mode.

For the temperature loss channel,
the input state of the bipartite system
is the product of the density matrices
given by
\begin{align}
\label{eq:rho}
\hat{\rho} = \hat{\rho}_{\ind{in}} \otimes \hat{\rho}_{\ind{th}},
\end{align}
where
$\hat{\rho}_{\ind{in}}$
is the density matrix of radiation 
\begin{align}
\label{eq:rho_in}
\hat{\rho}_{\ind{in}} =  \ket{\psi_{\ind{in}}} \bra{\psi_{\ind{in}}}
\end{align}
prepared in the pure state $\ket{\psi_{\ind{in}}}$,
whereas
the environment is in the thermal state 
\begin{align}
\label{eq:rho_th}
\rho_{\ind{th}} = \frac{1}{n_{\ind{th}}+1}
  \sum^{\infty}_{n=0}\e^{-n\beta\hbar\omega}\ket{n} \bra{n},
  \quad
n_{\ind{th}}=\Tr\{ \hcnj{\hat{b}}\hat{b}\hat{\rho}_{\ind{th}}\}= \frac{1}{\e^{\beta\hbar\omega}-1} .
\end{align}
where
$\beta = 1/(k_BT)$ is the inverse temperature parameter
($k_B$ is the Boltzmann constant and $T$ is the temperature),
$\hbar$ is the Planck constant, $\omega$ is the photon frequency,
$n_{\ind{th}}$ is the average  number of thermal photons (the mean thermal photon number).

Temporal evolution of the density matrix~\eqref{eq:rho}
is governed by the channel unitary~\eqref{eq:a_evolution}
as follows
\begin{align}
\label{eq:rho_tau}
  \hat{ \rho}(\tau)=\hat{U}_\tau \hat{ \rho} \hat{U}^{\dagger}_\tau.
\end{align}
We can now use Eqs.~\eqref{eq:a_evolution} and~\eqref{eq:rho_tau}
to obtain the normally ordered characteristic function
of the light as a function of the channel transmittance
\begin{align}
  &
  \label{eq:chi-tau}
  \chi(\alpha,\tau)=
  \Tr\{:\hat{D}(\alpha):\hat{\rho}(\tau)\}=
  \chi_{\ind{in}}(\sqrt{\tau}\alpha)\chi_{\ind{th}}(\sqrt{1-\tau}\alpha),
\end{align}
where $:\hat{D}(\alpha):=:\exp(\alpha\hcnj{\hat{a}}-\cnj{\alpha}\hat{a}):
=\exp(\alpha\hcnj{\hat{a}})\exp(-\cnj{\alpha}\hat{a})$
is the normally ordered displacement operator;
$\chi_{\ind{in}}$ and $\chi_{\ind{th}}$ are the input and thermal characteristic functions
given by
\begin{align}
  \label{eq:chi_in_th}
  \chi_{\ind{in}}(\alpha)= \Tr\{:\hat{D}(\alpha):\hat{\rho}_{\ind{in}}\},
  \quad
  \chi_{\ind{th}}(\alpha)= \Tr\{:\hat{D}(\alpha):\hat{\rho}_{\ind{th}}\}=\exp(-n_{\ind{th}}|\alpha|^2).
\end{align}
Note that the form of this result reproduces the characteristic function derived
by solving the thermal single-mode Lindblad equation
(see, e.g.,~\cite{Kiselev:entropy:2021})
so that  
the beam splitter transformation representation~\eqref{eq:a_evolution}
and the approach  based on the Lindblad dynamics
appear to be equivalent tools for modeling the temperature loss quantum channel.

Given the transmittance $\tau$,
it is not difficult to find
the average photon number of the output state $n_{\tau}$:
\begin{align}
  &
\label{eq:n_tau_trace}
  n_{\tau} =  \Tr\{ \hat{n} \hat{\rho}(\tau)\} =
  \Tr\{\hat{n}_\tau\hat{\rho}\}\equiv
  \avr{\hat{n}_\tau}=\tau n_{\ind{in}} + (1-\tau)
    n_{\ind{th}},
  \\
  &
    \label{eq:n_in}
    n_{\ind{in}}=
\Tr\{ \hat{n} \hat{\rho}_{\ind{in}}\} =
    \bra{\psi_{\ind{in}}} \hat{n} \ket{\psi_{\ind{in}}},
\end{align}
where
$\hat{n}=\hcnj{\hat{a}} \hat{a}$
and $  \hat{n}_{\tau}=\hcnj{\hat{a}}_\tau \hat{a}_\tau$.

Since $\hat{n}^2-\hat{n}=\hcnj{\hat{a}}\hcnj{\hat{a}} \hat{a}\hat{a}\equiv:\hat{n}^2:$,
the difference between the variance and the mean photon number 
for the quantum state $\hat{\rho}(\tau)$ can be computed as
the $q$-parameter given by
\begin{align}
  \label{eq:q_tau}
  q_\tau=\Tr\{: \hat{n}^2: \hat{\rho}(\tau)\} -n^2_\tau.
\end{align}
Clearly, for $\hat{\rho}(\tau)$,
the photon statistics is sub-Poissonian if and only if
the parameter~\eqref{eq:q_tau} is negative. 
Note that the ratio $q_\tau/n_\tau$ gives
the well-known Mandel $Q$-parameter~\cite{Mandel:bk:1995}, $Q^{(M)}_{\tau}$,
introduced in Refs.~\cite{Mandel:ol:1979,Mandel:prl:1983}.
Note that the $Q$-parameter is also
related to the normalized second-order correlation function
$g^{(2)}_{\tau}(0)=Q^{(M)}_{\tau}/n_\tau+1$ and
the Fano factor $F_{\tau}=Q^{(M)}_{\tau}+1$.

We can now use the relation
\begin{align}
    \label{eq:norm_n2_tau}
    \langle :\hat{n}^2_{\tau}: \rangle = 
  \tau^2
\avr{: \hat{n}^2:}
+
    2(1-\tau)^2 n_{\ind{th}}^2 +
    4 \tau(1-\tau)n_{\ind{in}}n_{\ind{th}}
\end{align}
derived with the help
of the identity
$\avr{\hcnj{\hat{b}}\hcnj{\hat{b}} \hat{b}\hat{b}} = 2 n_{\ind{th}}^2$
to deduce the explicit expression for the parameter~\eqref{eq:q_tau}
\begin{align}
    &
    \label{eq:q_tau-1}
    q_{\tau} = \langle :\hat{n}^2_{\tau}: \rangle -  n_{\tau}^2 = 
    \tau^2 q_{\ind{in}} + (1-\tau)^2 n_{\ind{th}}^2 +
    2 \tau(1-\tau)n_{\ind{in}}n_{\ind{th}},
\end{align}
where
\begin{align}
  \label{eq:m-in}
  q_{\ind{in}}=\Tr\{: \hat{n}^2: \hat{\rho}_{\ind{in}}\}-n_{\ind{ini}}^2=\avr{: \hat{n}^2:}-n_{\ind{in}}^2.
\end{align}
In the zero-temperature limit with $n_{\ind{th}}=0$,
formula~\eqref{eq:q_tau-1} shows that
losses cannot destroy the sub-Poissoinian statistics of input light.
The latter is generally no longer the case at non-zero temperatures.
In the subsequent section we will discuss the conditions
for negativity of the parameter $q_{\tau}$.

\section{Fluctuating losses and critical transmittance}
\label{sec:fluct}

Now we extend the results of the preceding section
to the case where the transmittance fluctuates
and thus should be treated as a random variable.
In this case, the output density matrix takes the generalized form: 
\begin{align}
\label{eq:rho_out}
  \hat{\rho}_{\ind{out}} =  \int_0^1 P(\tau)  \hat{ \rho}(\tau)  \mathrm{d}\tau,
\end{align}
where $P(\tau)$  is the probability density function (PDF) of the transmittance.
We can use
Eqs.~\eqref{eq:n_tau_trace} and~\eqref{eq:rho_tau}
with $\hat{\rho}(\tau)$ replaced by $\hat{\rho}_{\ind{out}}$
to introduce the output parameter $q_{\ind{out}}$
as follows
\begin{align}
\label{eq:q_out-1}
q_{\ind{out}} =\Tr\{:\hat{n}^2:\hat{\rho}_{\ind{out}}\}
  -\Tr\{\hat{n}\hat{\rho}_{\ind{out}}\}^2 = \langle  \langle :\hat{n}^2_{\tau}: \rangle \rangle_\tau - \langle n_{\tau}
                \rangle_\tau^2  = \langle q_\tau \rangle_\tau +
                [\langle  n_{\tau}^2\rangle_\tau - \langle n_{\tau} \rangle_\tau^2],
\end{align}
where $\avr{\cdots}_\tau=\int_0^1\cdots P(\tau)\dd\tau$.
After substituting formulas~\eqref{eq:n_tau_trace} and~\eqref{eq:q_tau}
into Eq.~\eqref{eq:q_out-1},
we obtain the parameter $q_{\ind{out}}$ in the following explicit form:
\begin{align}
  &
\label{eq:q_out-2}
  q_{\ind{out}} = \overline{\tau}^2  (q_{\ind{in}} - n_{\ind{in}}^2) +
  (\overline{\tau} n_{\ind{in}} +(1-\overline{\tau})n_{\ind{th}})^2 +
  \notag
  \\
  &
  \mathrm{Var}(\tau ) \{q_{\ind{in}} - n_{\ind{in}}^2 + 2(n_{\ind{in}}-n_{\ind{th}})^2)\},
\end{align}
where  $\overline{\tau}=\avr{\tau}_\tau$ is the averaged transmittance and 
$\mathrm{Var}(\tau )=\avr{\tau^2}_\tau-\overline{\tau}^2$ is the variance of the transmittance.

Since the transmittance varies from zero to unity, $0\le \tau\le 1$,
and the variance $\mathrm{Var}(\tau )$ cannot be negative,
$\mathrm{Var}(\tau )\ge 0$,
the mean of squared transmittance cannot exceed the average transmittance
and must satisfy the condition:
$\overline{\tau}^2\le\avr{\tau^2}_\tau\le\overline{\tau}$.
As a result, the variance
is bounded from above by the product
$(1 - \overline{\tau})\overline{\tau}$:
$ 0\le\mathrm{Var}(\tau )\le (1 - \overline{\tau})\overline{\tau}$.
In our model, we
assume that
$\avr{\tau^2}_\tau=F \overline{\tau}+(1-F)\overline{\tau}^2$,
so that the variance takes the following form 
 \begin{align}  
 \label{eq:var_tau_parametr}
 \mathrm{Var}(\tau )=\avr{\tau^2}-\overline{\tau}^2= F (1 - \overline{\tau})\overline{\tau}
\end{align}
where $0\le F \le 1$ is the parameter  characterizing  
the strength of fluctuations.

In what follows, we shall treat the fluctuation strength $F$
as a phenomenological  parameter which is independent of $\overline{\tau}$.
The variance representation~\eqref{eq:var_tau_parametr} can now be used
to express the $q$-parameter~\eqref{eq:q_out-2}
in terms of $\overline{\tau}$ as follows
\begin{align}
  &
    \label{eq:q_out-3}
    q_{\ind{out}}(\overline{\tau}) = \overline{\tau}^2 ( q_{\ind{in}} -a ) + \overline{\tau} (a -
    n_{\ind{th}}^2) + n_{\ind{th}}^2,
\end{align}
where
\begin{align}
  \label{eq:a-q}
    a=2 n_{\ind{in}}n_{\ind{th}}-n^2_{\ind{th}} + g F,
    \quad
    g =  q_{\ind{in}} - n^2_{\ind{in}} + 2(n_{\ind{in}} - n_{\ind{th}} )^2.
\end{align}

In the limiting case of
transparent medium described by  the identity channel with $\overline{\tau}=1$,
the $q$-parameter is equal to its initial value
$q_{\ind{in}}$ given by Eq.~\eqref{eq:m-in}.
For sub-Poissonian light,
the parameter $q_{\ind{in}}$ is negative, so that $q_{\ind{out}}(1)= q_{\ind{in}}< 0 $.
In the opposite case with vanishing transmittance, $\overline{\tau}=1$,
from Eq.~\eqref{eq:q_out-3}, the $q$-parameter is positive,
$q_{\ind{out}}(0) = n_{\ind{th}}^2>0$.
Clearly, this implies that
the transmitted light will remain sub-Poissonian
provided the average transmittance exceeds its critical value:
$\overline{\tau}>\tau_c$.

From Eq.~\eqref{eq:q_out-3}, it is not difficult to
find the analytical expression for the critical transmittance
(the transmittance threshold).
So, we deduce the inequality
\begin{align}
\label{eq:tau_с}
\overline{\tau} > \tau_c = -\frac{(a-n_{\ind{th}}^2) + \sqrt{(a+n_{\ind{th}}^2)^2 - 4n_{\ind{th}}^2q_{\ind{in}}}}{2(q_{\ind{in}}-a)}
\end{align}
giving the condition that
effects of transmittance fluctuations and temperature
fail to destroy the sub-Poissonian statistics of light and
the $q$-parameter of the transmitted (output) light is negative,
$q_{\ind{out}}<0$.
 
We conclude this section with the remark on two important special cases
with $T=0$ and $F=0$, respectively.
In the zero-temperature  limit,
the expression of the transmittance threshold is simplified as follows
\begin{align}
\label{eq:tau_c-Tzero}
\tau_c|_{T=0} = \frac{F (q_{\ind{in}} + n_{\ind{in}}^2)}{F (q_{\ind{in}} + n_{\ind{in}}^2) - q_{\ind{in}}},
\end{align}
whereas
the critical transmittance for
the case, where fluctuations of the transmittance are negligible, is given by
\begin{align}
\label{eq:tau_c-Fzero}
\tau_c|_{F=0} = \frac{n_{\ind{th}}}{n_{\ind{th}}+\sqrt{n_{\ind{in}}^2-q_{\ind{in}}} - n_{\ind{in}}}.
\end{align}
Clearly, when $F$ and $T$ are both vanishing, the critical transmittance is zero.
This is the well-known case of non-fluctuating pure-loss bosonic channel
where $q_{\ind{out}}=\tau^2 q_{\ind{in}}$ (see Eq.~\eqref{eq:q_tau-1} with $n_{\ind{th}}=0$).

\begin{figure*}[!htb]
  \centering
  \begin{subfigure}{0.45\textwidth}
    \includegraphics[width=\linewidth]{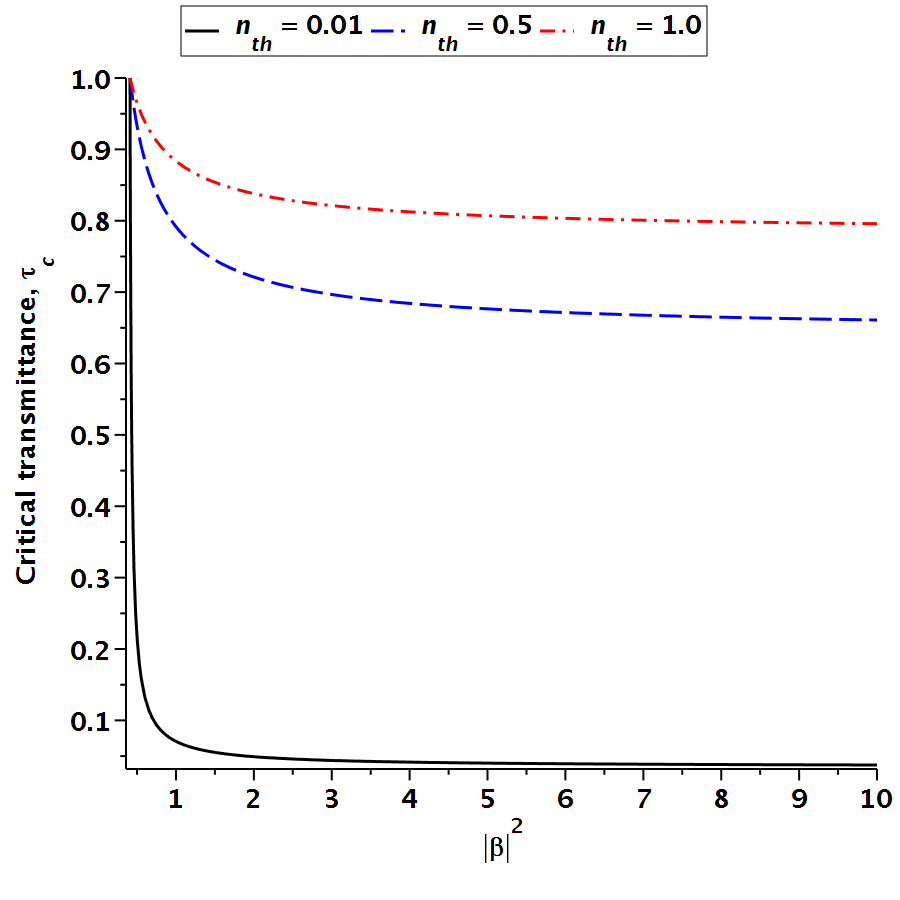}
    \caption{$F=0.0$}
    \label{fig:sq-Fa}
\end{subfigure}
  \begin{subfigure}{0.45\textwidth}
    \includegraphics[width=\linewidth]{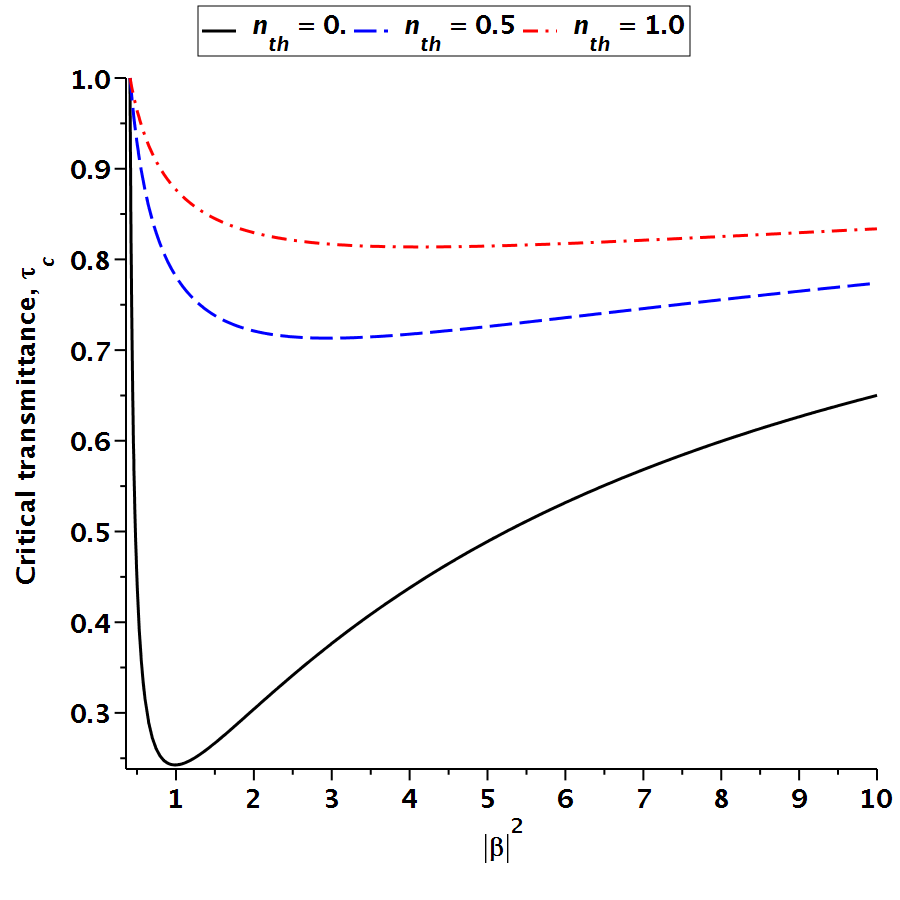}
    \caption{$F=0.1$}
    \label{fig:sq-Fb}
\end{subfigure}
  \begin{subfigure}{0.45\textwidth}
    \includegraphics[width=\linewidth]{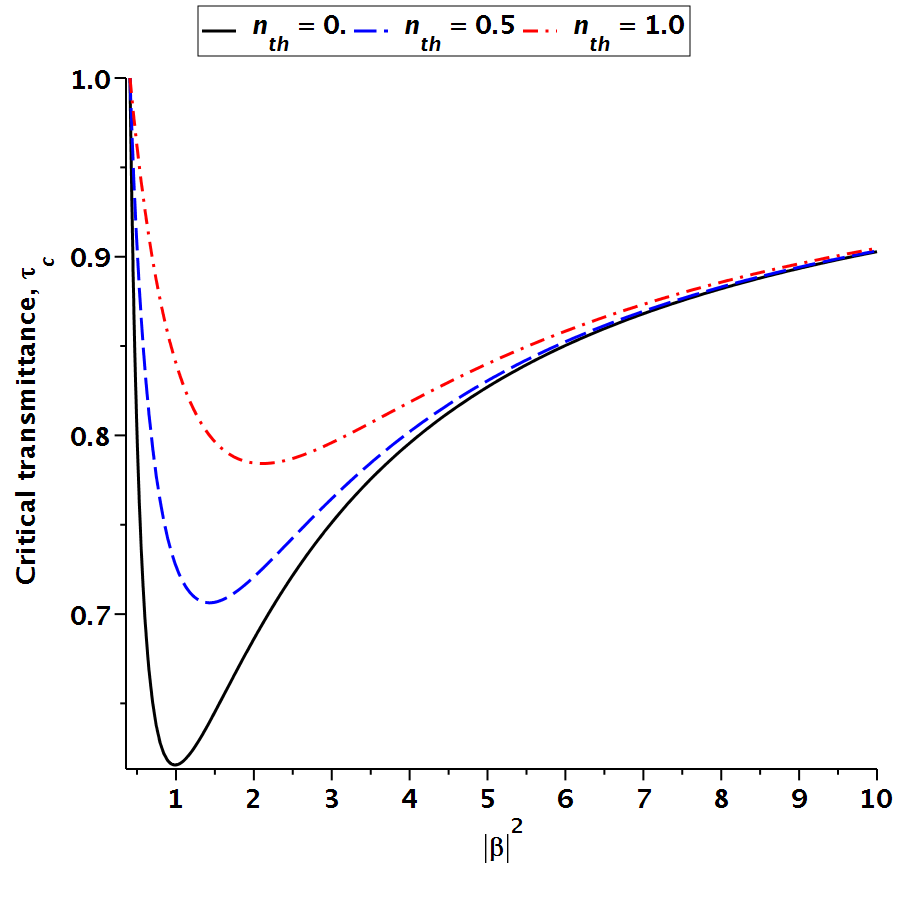}
    \caption{$F=0.5$}
    \label{fig:sq-Fc}
\end{subfigure}
  \begin{subfigure}{0.45\textwidth}
    \includegraphics[width=\linewidth]{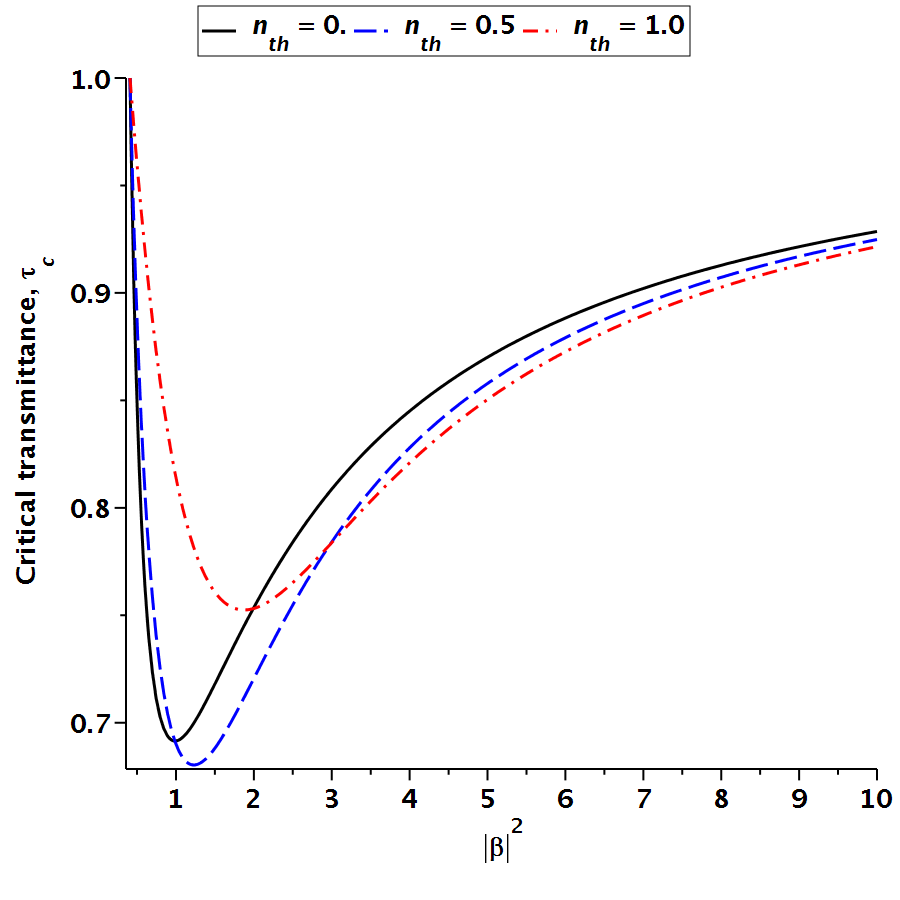}
    \caption{$F=0.7$}
    \label{fig:sq-Fd}
\end{subfigure}
  \caption{Dependence of the critical transmittance  $\tau_c$  on the squared amplitude of
    the displacement, $|\beta|^2$,
    for the displaced squeezed state~\eqref{eq:sq-state}
    computed
at different values of the fluctuation strength parameter $F$
    for $r=0.4$ and $\phi=\pi/2$.
 }
  \label{fig:sq-F}
\end{figure*}

\section{Effects of fluctuations and temperature}
\label{sec:temp}

In this section, we apply our analytical results
to the two special cases of
input states: the displaced squeezed states
and the  odd Schr\"odinger cat states.
More specifically,
we shall study how the temperature and transmitance fluctuations
influence the transmittance threshold for these states. 

\subsection{Squeezed light}
\label{subsec:sq}

The density matrix 
\begin{align}
  &
    \label{eq:sq-state}
    \hat{\rho}_{\ind{in}} =
    \ket{\psi_{\ind{sq}}} \bra{\psi_{\ind{sq}}} , \quad
    \ket{\psi_{\ind{sq}}} =  \hat{D}(\beta) \hat{S}(\xi) | 0 \rangle
\end{align}
describes the case of displaced squeezed vacuum states
expressed using
the squeezing operator, 
$\hat{S}(\xi)$, and the displacement operator,
$\hat{D}(\beta)$, given by
\begin{align}
  &
    \label{eq:SD-operators}
  \hat{S}(\xi)= e^{(\xi(\Ah)^2-\xi^* \ah^2)/2},
\quad
  \hat{D}(\beta)=e^{\beta \Ah -\beta^* \ah},
\end{align}
$\xi=r \e^{i\psi}$ is the squeezing parameter 
  and $\beta=|\beta| \e^{i\theta}$ is the displacement.
For the squeezed state~\eqref{eq:sq-state},
it is rather straightforward to derive
the expressions for 
the parameters $n_{\ind{in}}=n_{\ind{sq}}$ and
$q_{\ind{in}}=q_{\ind{sq}}$ 
that enter the expression for the critical transmittance~\eqref{eq:tau_с}: 
\begin{align}
  &
    \label{eq:nm-sq}
n_{\ind{sq}} = |\beta|^2+\sinh^2 r,\quad
    q_{\ind{sq}} = q_2 |\beta|^2+\sinh^2r \cosh 2r,
\end{align}
where
$q_2=2\sinh^2r +\sinh 2r\cos(2\phi)$ and $\phi=\theta-\psi/2$.

From formula~\eqref{eq:nm-sq},
the case where 
the photon statistics of the squeezed light
is sub-Poissonian with negative 
parameter $q_{\ind{sq}}$
may occur only
when the coefficient $q_2$ is negative.
The latter requires the angle $\phi$ and the squeezing parameter $r$
to meet the inequality
\begin{align}
  \label{eq:cond-sq}
  |\tan\phi|> e^r,
\end{align}
so that the squeezed light will be sub-Poissonian
only when the squared displacement amplitude
is above its critical value:
\begin{align}
&
  \label{eq:beta-sq}
  |\beta|^2 > \frac{e^r \sinh r \cosh 2 r}{2(\sin^2\phi-e^{2 r}\cos^2\phi)}\equiv \beta_c^2.
\end{align}
Clearly,
when $\cos\phi=0$,
the condition~\eqref{eq:cond-sq} is always fulfilled
and $\beta_c$ takes the minimal value that grows with $r$.

\begin{figure*}[!htb]
  \centering
  \begin{subfigure}{0.45\textwidth}
    \includegraphics[width=\linewidth]{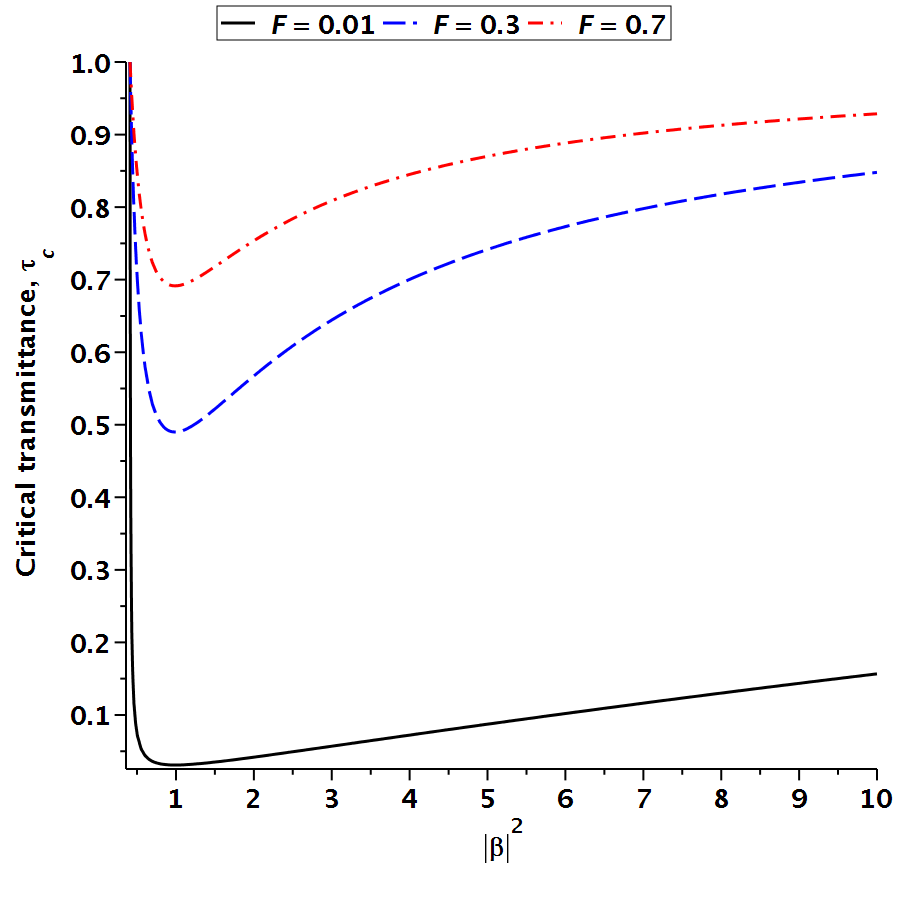}
    \caption{$n_{\ind{th}}=0.0$}
    \label{fig:sq-ntha}
\end{subfigure}
  \begin{subfigure}{0.45\textwidth}
    \includegraphics[width=\linewidth]{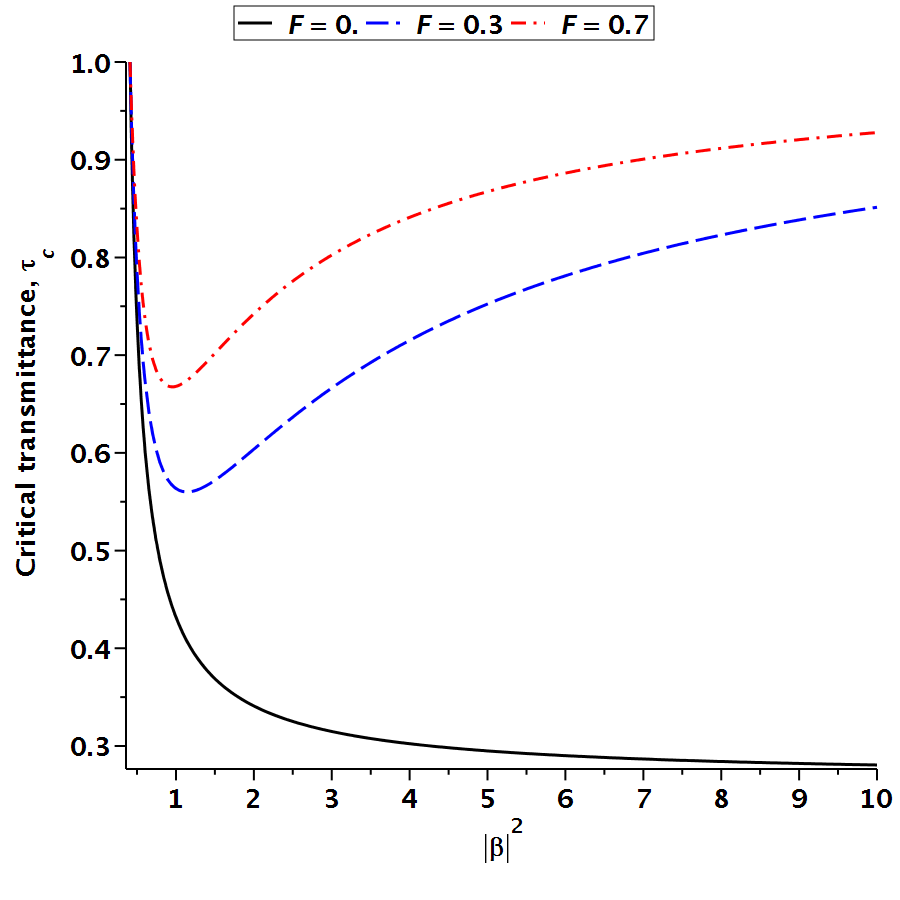}
    \caption{$n_{\ind{th}}=0.1$}
    \label{fig:sq-nthb}
\end{subfigure}
  \begin{subfigure}{0.45\textwidth}
    \includegraphics[width=\linewidth]{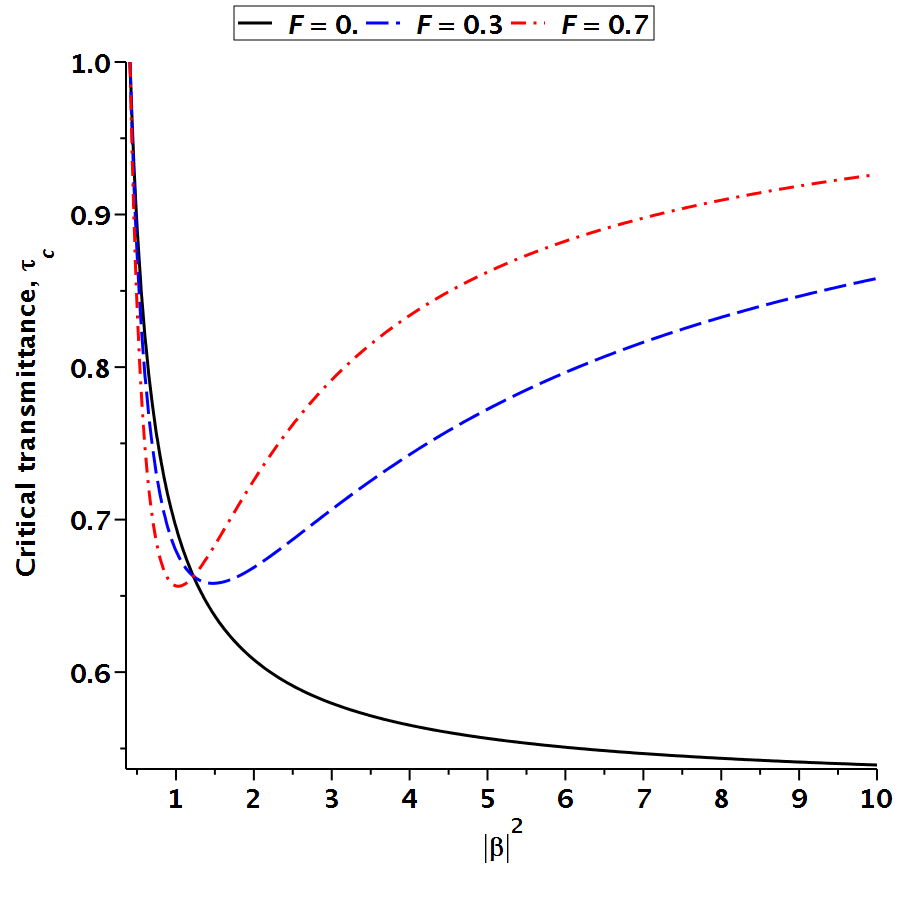}
    \caption{$n_{\ind{th}}=0.3$}
    \label{fig:sq-nthc}
\end{subfigure}
  \begin{subfigure}{0.45\textwidth}
    \includegraphics[width=\linewidth]{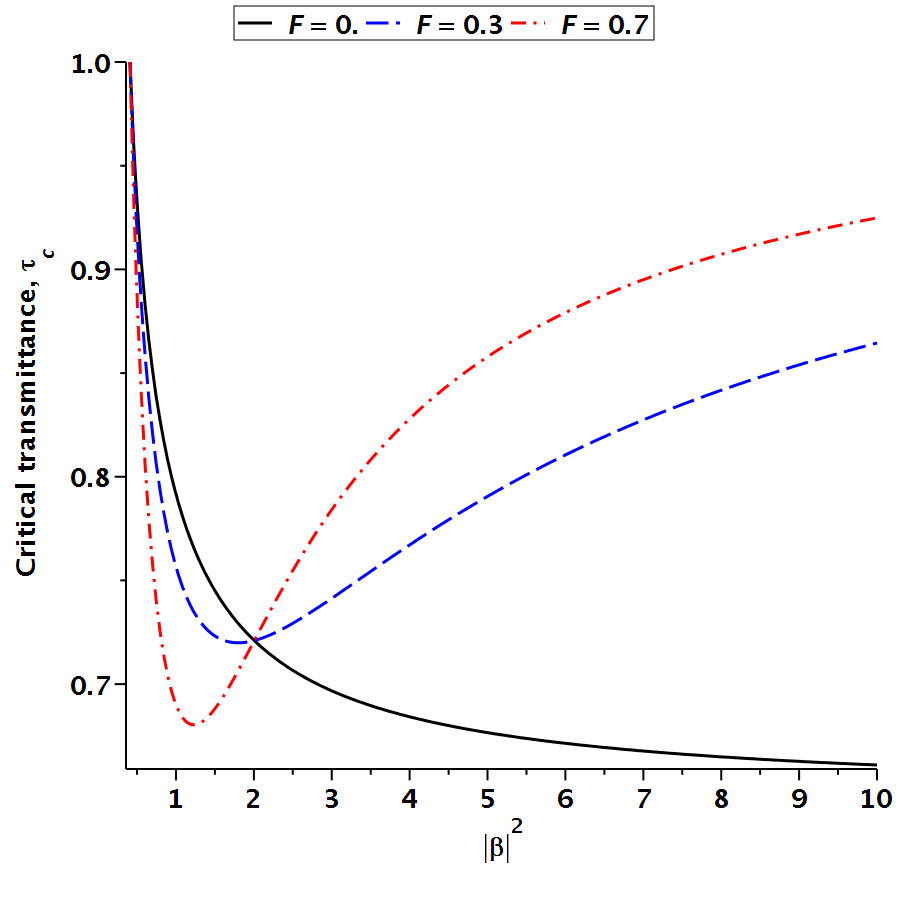}
    \caption{$n_{\ind{th}}=0.5$}
    \label{fig:sq-nthd}
\end{subfigure}
  \caption{Dependence of the critical transmittance  $\tau_c$  on the squared amplitude of
    the displacement, $|\beta|^2$,
    for the displaced squeezed state~\eqref{eq:sq-state}
    with $r=0.4$ and $\phi=\pi/2$
    computed
at different values of the mean thermal photon number $n_{\ind{th}}$.
 }
  \label{fig:sq-nth}
\end{figure*}

In Figs.~\ref{fig:sq-F} and~\ref{fig:sq-nth},
the threshold transmittance is plotted against the squared displacement amplitude,
$|\beta|^2$, at $\phi=\pi/2$.
The numerical results for $\tau_c$ are evaluated using
Eq.~\eqref{eq:tau_с} with
formula~\eqref{eq:nm-sq} giving
the mean photon number 
$n_{\ind{in}}$ and the $q$-parameter
$q_{\ind{in}}$.

At $|\beta|=\beta_c$,
the $q$-parameter vanishes and thus the threshold transmittance equals unity.
So, in the close vicinity of  $\beta_c$,
the transmittance $\tau_c$ drops as
$|\beta|^2$ increases.

From Eq.~\eqref{eq:tau_c-Fzero},
it is not difficult to show that, when fluctuations are negligible,
$\tau_c$ monotonically decreases with $|\beta|^2$
approaching the value
$n_{\ind{th}}/(n_{\ind{th}}-q_2/2)$ (see Fig.~\ref{fig:sq-Fa}).
By contrast,
from the relation~\eqref{eq:tau_c-Tzero}
describing the zero-temperature case,
in the large displacement limit,
the critical transmittance approaches unity.
As a result, the $|\beta|^2$-dependence of $\tau_c$
reveals nonmonotonic behaviour
illustrated in Fig.~\ref{fig:sq-ntha}.
It can also be seen that such behaviour where
the critical transmittance exhibits
a local minimum occurs at non-vanishing temperatures
provided $F\ne 0$.

\begin{figure*}[!htb]
  \centering
  \begin{subfigure}{0.45\textwidth}
    \includegraphics[width=\linewidth]{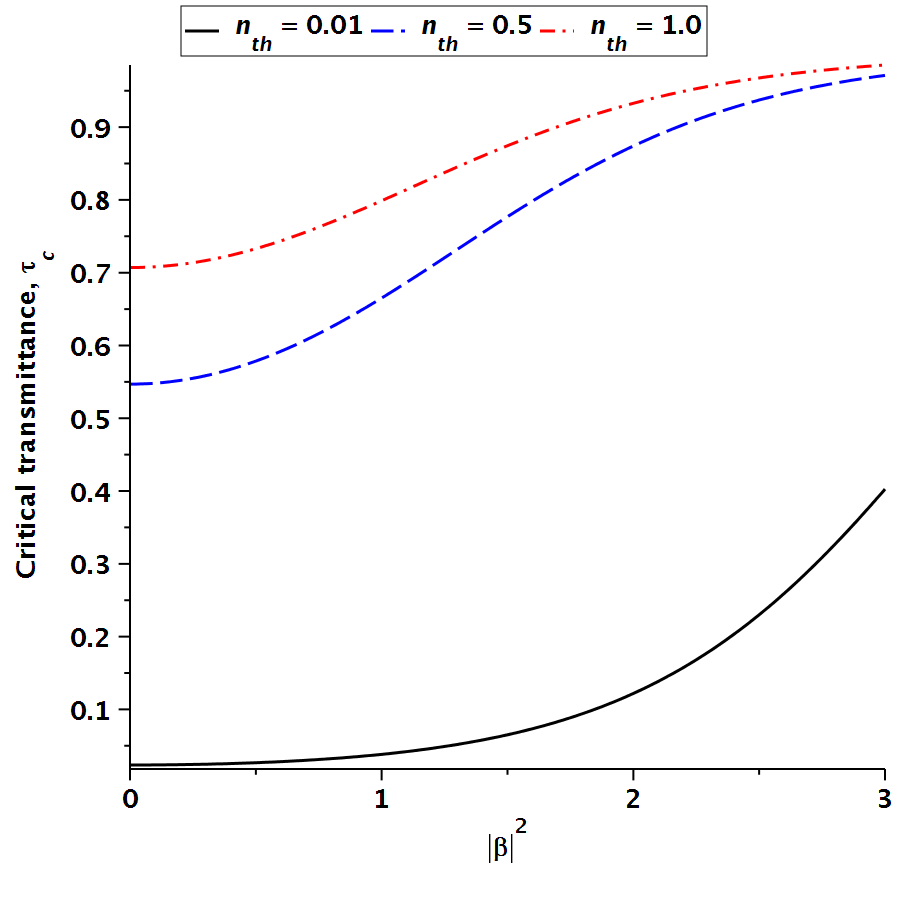}
    \caption{$F=0.0$}
     \label{fig:cat-Fa}
\end{subfigure}
  \begin{subfigure}{0.45\textwidth}
    \includegraphics[width=\linewidth]{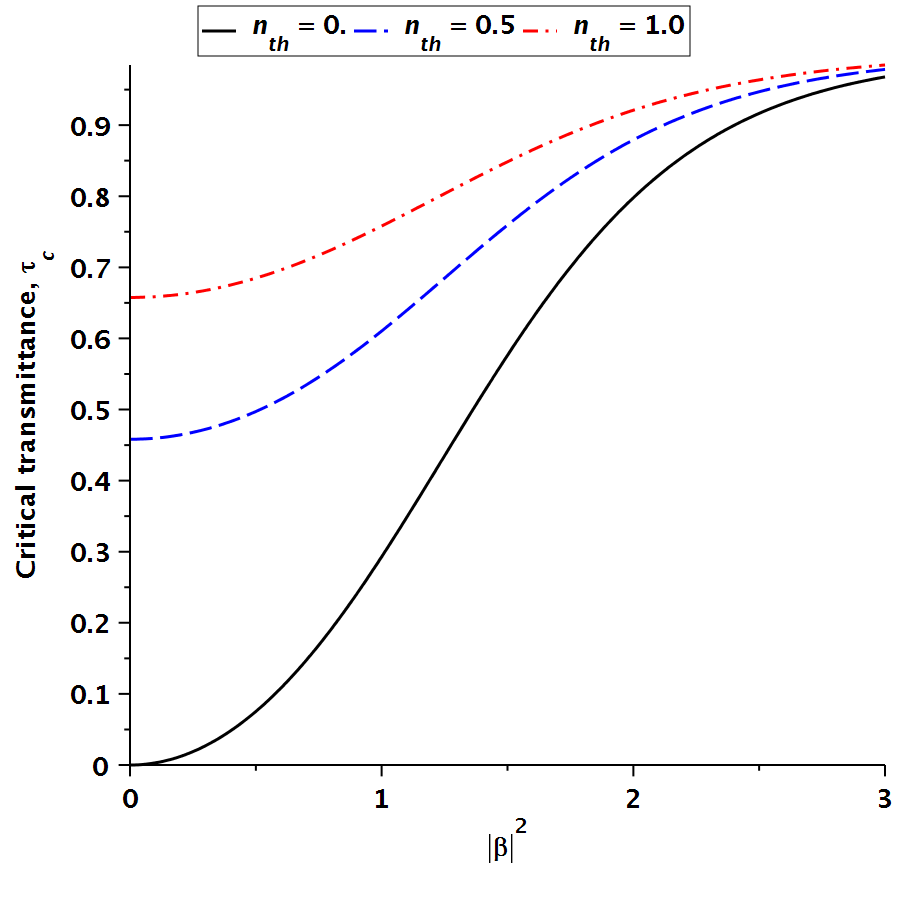}
    \caption{$F=0.3$}
     \label{fig:cat-Fb}
\end{subfigure}
  \begin{subfigure}{0.45\textwidth}
    \includegraphics[width=\linewidth]{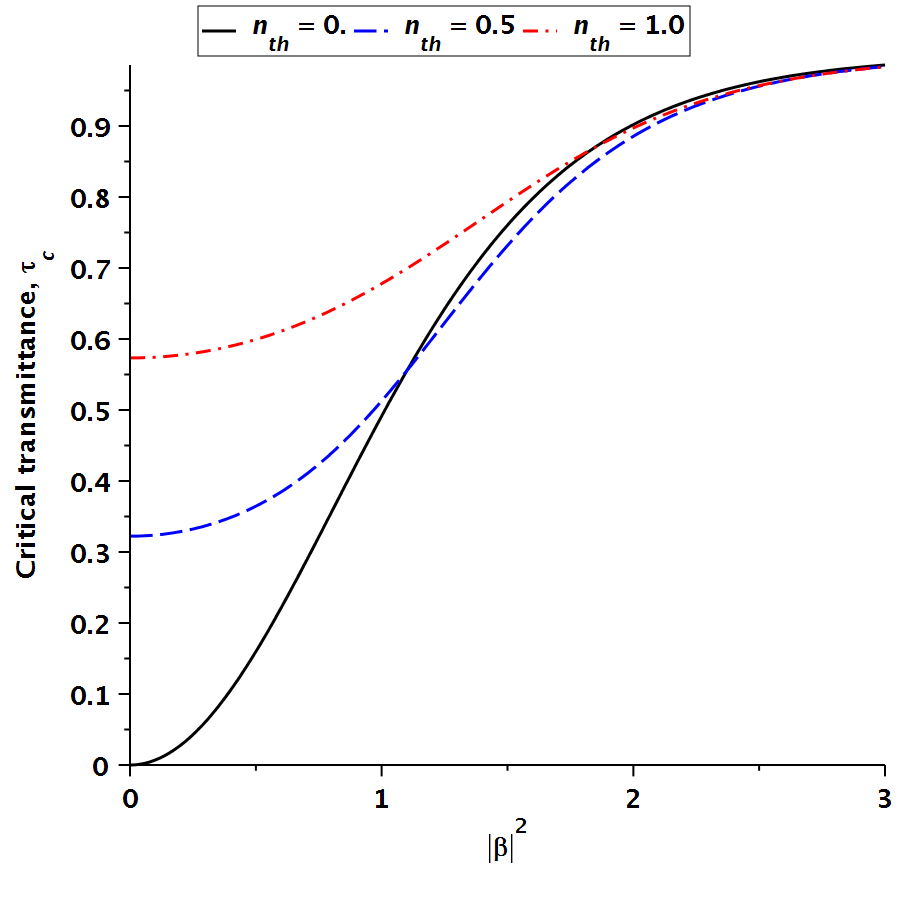}
    \caption{$F=0.7$}
     \label{fig:cat-Fc}
\end{subfigure}
  \begin{subfigure}{0.45\textwidth}
    \includegraphics[width=\linewidth]{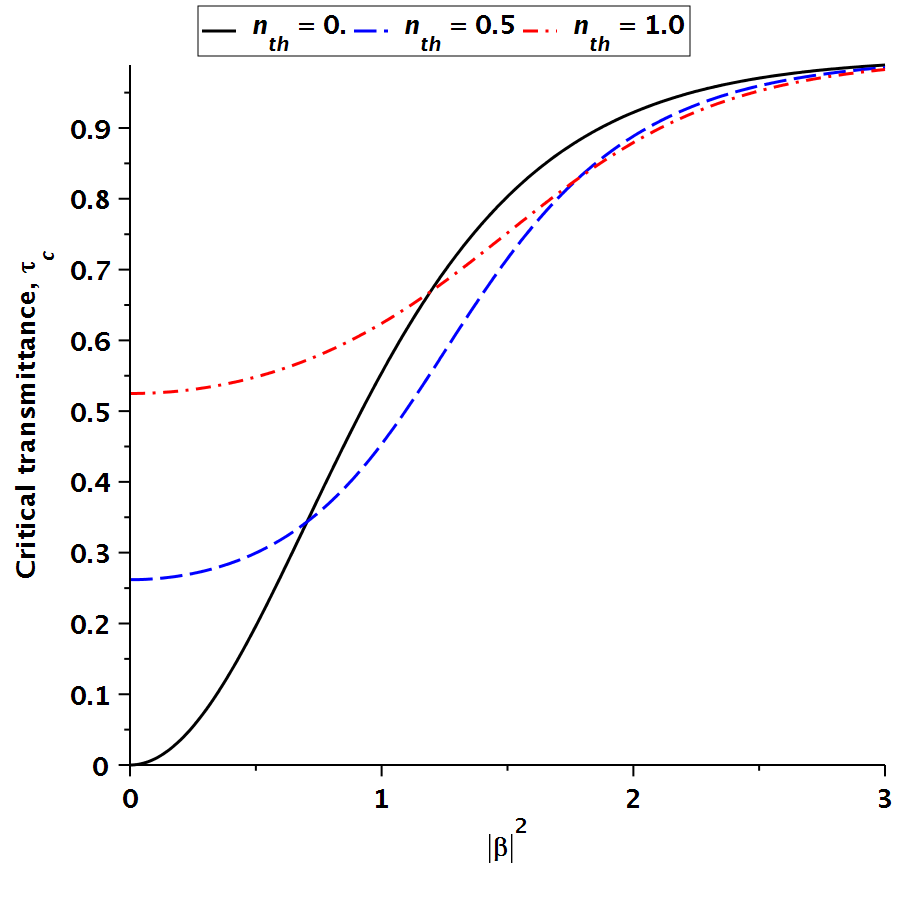}
    \caption{$F=0.8$}
     \label{fig:cat-Fd}
\end{subfigure}
  \caption{Dependence of the critical transmittance  $\tau_c$  on the squared amplitude of
    the displacement, $|\beta|^2$,
    for the odd cat state~\eqref{eq:cat-state}
    computed
at different values of the fluctuation strength parameter $F$.
 }
  \label{fig:cat-F}
\end{figure*}

Normally, it is expected
that the threshold will get higher at elevated temperatures.
This is the case for the curves shown in
Figs.~\ref{fig:sq-Fa}--\ref{fig:sq-Fc}.

Interestingly,
in the case of intense fluctuations
with sufficiently high values of fluctuation strength,
the latter is no longer the case.
From Figure~\ref{fig:sq-Fd} which shows
the $\tau_c$-vs-$|\beta|^2$ curves computed at $F=0.7$,
it is seen that the long displacement part of
the zero-temperature curve appears to be above
the curves with non-zero $n_{\ind{th}}$.

Similarly, the threshold is expected to increase with the fluctuation strength
$F$.
It is not difficult to show that the threshold $\tau_c$
grows with $F$ only if the coefficient
$g$ given by Eq.~\eqref{eq:a-q} is positive.
According to formula~\eqref{eq:a-q}, this coefficient
quadratically depends on $n_{\ind{th}}$ and
it takes negative values within the interval:
\begin{align}
  \label{eq:n_pm}
  n_{-}<n_{\ind{th}}<n_{+},\quad
  n_{\pm}=n_{\ind{in}}\pm\sqrt{\frac{n_{\ind{in}}^2-q_{\ind{in}}}{2}}.
\end{align}
So, in the low temperature region where $n_{\ind{th}}<n_{-}$
the critical transmittance is an increasing function of the fluctuation strength.
The curves presented in Figs.~\ref{fig:sq-ntha}--\ref{fig:sq-nthb}
provide support to this conclusion.

For the cases of elevated temperatures
illustrated in Figs.~\ref{fig:sq-nthc}--\ref{fig:sq-nthd},
the short displacement part of the curves
where $n_{\ind{th}}$ meets the condition~\eqref{eq:n_pm} is arranged differently.
In this part,
the curve with negligibly small fluctuations determines
the largest value of $\tau_c$.
Note that all the curves shown in Figs.~\ref{fig:sq-nthc}--\ref{fig:sq-nthd}
intersect at the point where
the coefficient $g$ vanishes with $n_{\ind{th}}=n_{-}$.

\begin{figure*}[!htb]
  \centering
  \begin{subfigure}{0.45\textwidth}
    \includegraphics[width=\linewidth]{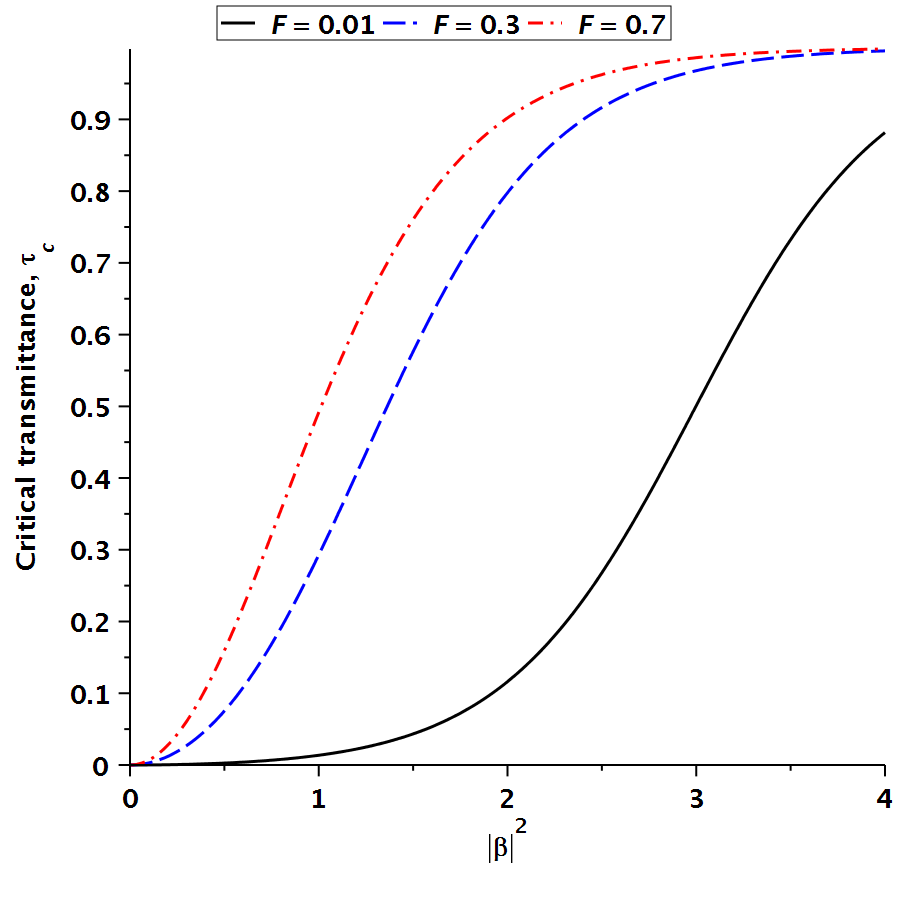}
    \caption{$n_{\ind{th}}=0.0$}
      \label{fig:cat-ntha}
\end{subfigure}
  \begin{subfigure}{0.45\textwidth}
    \includegraphics[width=\linewidth]{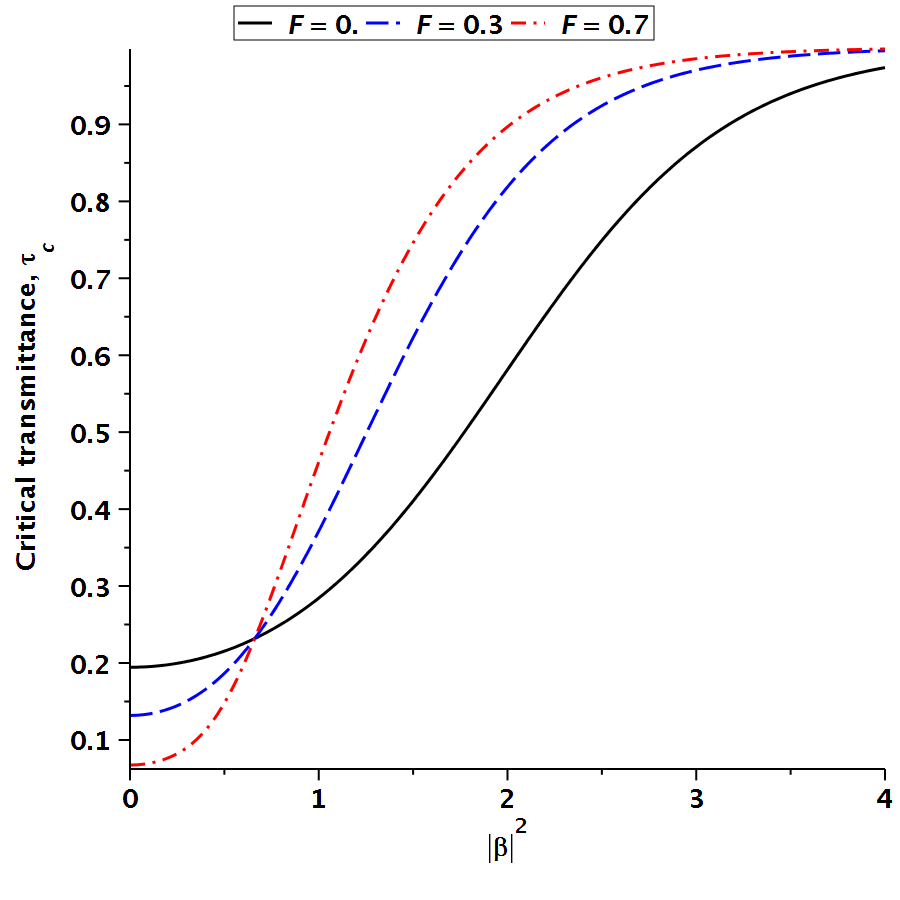}
    \caption{$n_{\ind{th}}=0.1$}
      \label{fig:cat-nthb}
\end{subfigure}
  \begin{subfigure}{0.45\textwidth}
    \includegraphics[width=\linewidth]{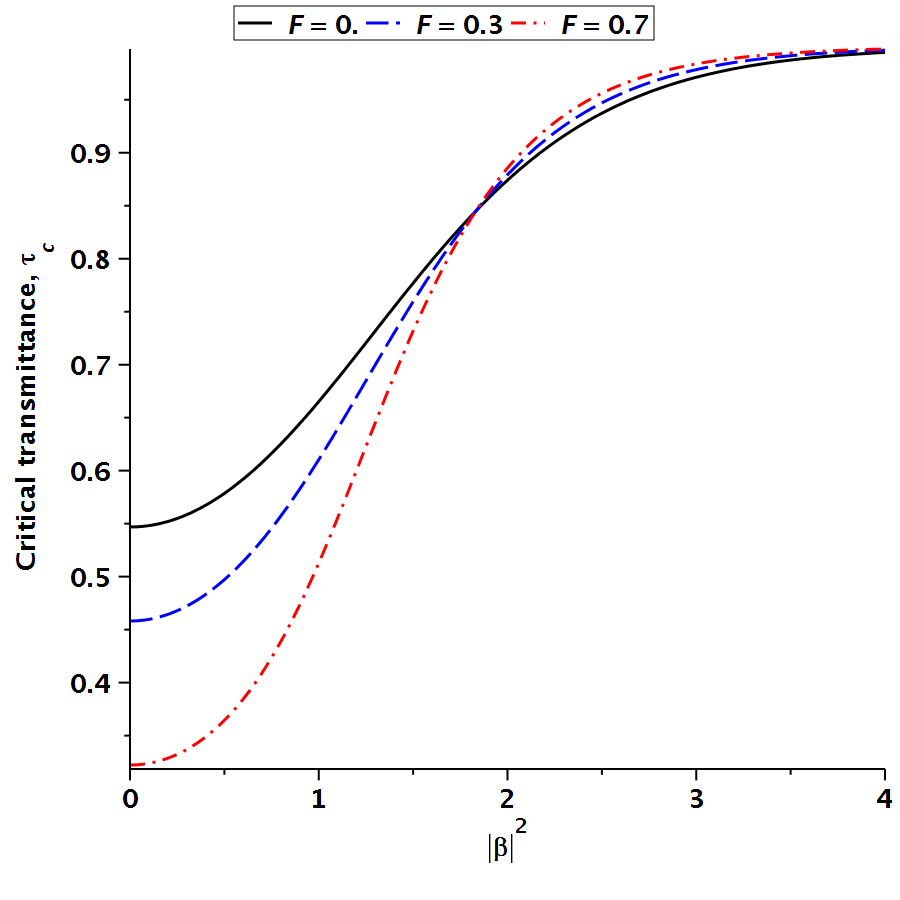}
    \caption{$n_{\ind{th}}=0.5$}
      \label{fig:cat-nthc}
\end{subfigure}
  \begin{subfigure}{0.45\textwidth}
    \includegraphics[width=\linewidth]{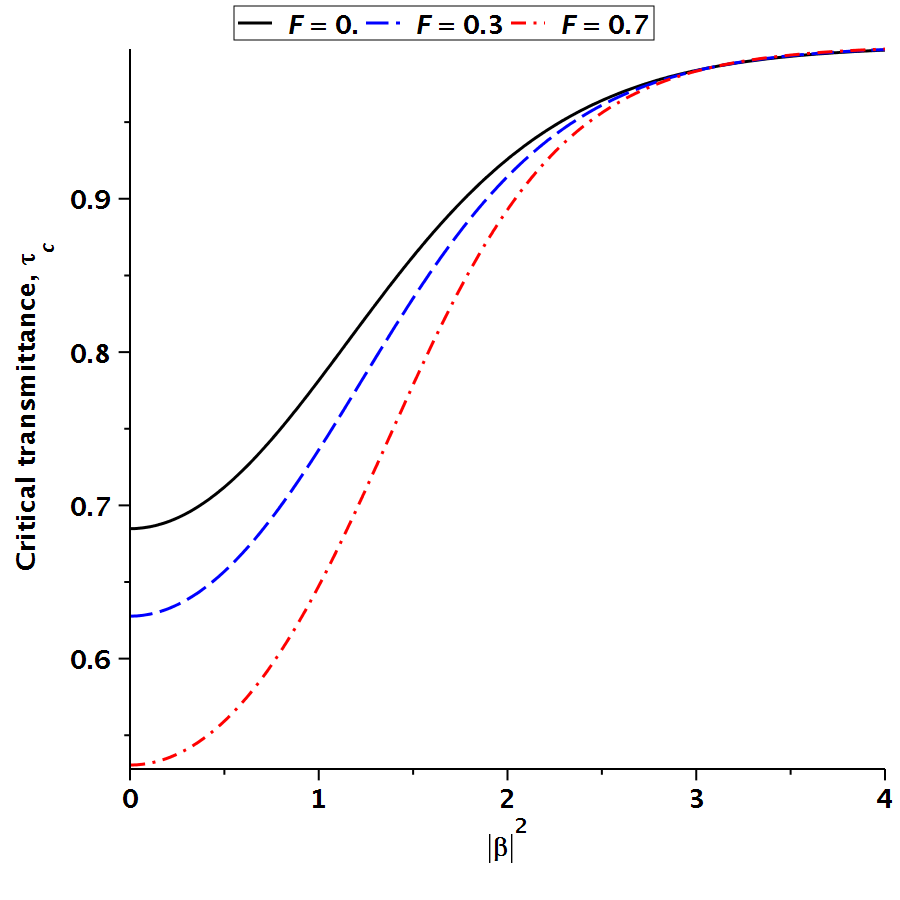}
    \caption{$n_{\ind{th}}=0.9$}
      \label{fig:cat-nthd}
\end{subfigure}
  \caption{Dependence of the critical transmittance  $\tau_c$  on the squared amplitude of
    the displacement, $|\beta|^2$,
    for the odd cat state (see Eq.~\eqref{eq:cat-state})
    computed at different values of the mean thermal photon number $n_{\ind{th}}$.
 }
  \label{fig:cat-nth}
\end{figure*}

\subsection{Odd optical cats}
\label{subsec:cats}

Similar to the case of squeezed states,
formulas
\begin{align}
&
  \label{eq:cat-state}
  \hat{\rho}_{\ind{in}} = \ket{\psi_{\ind{cat}}} \bra{\psi_{\ind{cat}}} , \quad
  \ket{\psi_{\ind{cat}}} = \frac{1}{\sqrt{2(1- \e^{-2|\beta|^2})}} (\ket{\beta}-\ket{-\beta}),
  \\
&
n_{\ind{cat}} = |\tilde{\beta}|^2 \cosh|\beta|^2, \quad q_{\ind{cat}} = -|\tilde{\beta}|^4, 
\end{align}
where $\displaystyle|\tilde{\beta}|^2 = \frac{|\beta|^2}{\sinh{|\beta|^2}}$, 
present the analytical results needed to evaluate the critical transmittance for
the odd (antisymmetric) cat states, $\ket{\psi_{\ind{cat}}}$.

In Fig.~\ref{fig:cat-F} we show how
the fluctuation strength affects
the $|\beta|^2$-dependencies of the critical transmittance
at different temperatures.
From Eq.~\eqref{eq:tau_c-Fzero},
in the weak fluctuation limit with $F=0$
(see Fig.~\ref{fig:cat-Fa}),
the starting value of $\tau_c$
is $n_{\ind{th}}/(n_{\ind{th}}+\sqrt{2}-1)$ and it
monotonically increases approaching unity.
By contrast to the case of the squeezed states,
for the odd optical cats,
all the curves shown in Figs.~\ref{fig:cat-F} and~\ref{fig:cat-nth}
reveal similar behaviour.
Similar to the squeezed states,
it turned out that,
at sufficiently strong fluctuations (sufficiently large fluctuation strength $F$)
and sufficiently large $|\beta|^2$,
the zero-temperature curve gives
the largest value of $\tau_c$.

Figure~\ref{fig:cat-nth} presents the graphs
illustrating the temperature induced effects.
Since,
in the limiting case of single photon state with
$\beta=0$,
the mean photon number $n_{\ind{cat}}$ is unity
and $q_{\ind{cat}}=-1$,
all the zero-temperature curves described by Eq.~\eqref{eq:tau_c-Tzero}
start from zero (see Fig.~\ref{fig:cat-ntha}).
As is depicted in Figs.~\ref{fig:cat-nthb} and~\ref{fig:cat-nthc},
at non-vanishing temperatures
and sufficiently small $|\beta|$,
the values of $\tau_c$ are
dominated by the zero-fluctuation curve with $F=0$
(for the single photon state, the zero-fluctuation value of $\tau_c$
is $n_{\ind{th}}/(n_{\ind{th}}+\sqrt{2}-1)$).
Referring to Fig.~\ref{fig:cat-nthd},
in the high temperature region with $n_{\ind{th}}\ge 0.9$,
the latter holds for all values of $|\beta|$.

Since in the zero amplitude limit with $\beta=0$,
the odd state~\eqref{eq:cat-state}
becomes the single-photon state $\ket{1}$, it is instructive
to briefly discuss the case of the Fock states $\ket{n}$.
For such states, we have
$m_{\ind{in}}=-n$ and $n_{\ind{in}}=n$.
So, at low temperatures with $n_{\ind{th}}<n_{-}=n-\sqrt{n(n+1)/2}$,
an increase in $F$ will enhance the threshold.
Interestingly, in the special case with $n=1$,
we have  $n_{-}=0$.
As a result, all the curves in Fig.~\ref{fig:cat-ntha} start from the zero
and, in Figs.~\ref{fig:cat-nthb}--\ref{fig:cat-nthd},
the thresholds at $\beta=0$ are below the zero-fluctuation value of $\tau_c$:
$n_{\ind{th}}/(n_{\ind{th}}+\sqrt{2}-1)$.

\section{Conclusions}
\label{sec:discussion}

In this paper,
we have studied effects of the thermal-loss channel with fluctuating transmittance
on the sub-Poissonian light whose
non-classicality is characterized by
the $q$-parameter (see Eq.~\eqref{eq:q_tau}).
We have combined the input-output relation for the $q$-parameter~\eqref{eq:q_out-2}
with the variance of the transmittance
parameterized using the fluctuation strength parameter~\eqref{eq:var_tau_parametr}
to show that the condition for sub-Poissonian statistics of photon at the channel output
is determined by the critical transmittance~\eqref{eq:tau_с}.
For the cases of the
displaced squeezed state (see Eq.~\eqref{eq:sq-state})
and the odd optical cat state (see Eq.~\eqref{eq:cat-state}),
the critical transmittance is computed as a function of
the squared displacement amplitude, $|\beta|^2$,
at different values of temperature and the fluctuation parameter.
In contrast to what is expected, under certain conditions,
an increase in either the fluctuation strength
or the temperature may result in a decrease in
the critical transmittance.

Note that the key point greatly simplifying our analysis
is the parameterization of the transmittance
variance~\eqref{eq:var_tau_parametr},
where the fluctuation strength $F$
is introduced as a phenomenological parameter
which is independent of the mean transmittance
$\overline{\tau}$.
A more sophisticated treatment of
the atmospheric channels~\cite{Vasylyev:prl:2016,Vasylyev:pra:2018,Klen:pra:2023}
requires computing the first-order and second-order moments
of the transmittance, $\avr{\tau}=\overline{\tau^2}$ and $\avr{\tau^2}=\overline{\tau^2}$,
from the correlation functions
derived using the phase approximation of
the Huygens-Kirchhoff method~\cite{Banakh:ol:1977,Banakh:ol:1979}.

\section*{Acknowledgements}
 This work  was supported by the Russian Science Foundation (project No. 24–29–00786).




%

\end{document}